# DESIGN AND CONSTRUCTION OF A MICROCONTROLLER BASED ELECTRONIC MOVING MESSAGE DISPLAY

[1]Ogunseye Titus T., [2]Egbeyale Godwin B., [2] [3]Bello A. K. and [4]Ajani S.A.

[1]University of Ibadan, Oyo State, Nigeria

[2,4]Kwara State University, Malete, Kwara State, Nigeria

Bell University, Ota, Ogun State, Nigeria

Corresponding authors: tseyetaofik@yahoo.com and godwinegbeyale@gmail.com

## Abstract

This work presents a simple design and implementation of a microcontroller-based electronic moving message display system. The design involves the arrangement of Light Emitting Diodes (LEDs) and the programming of a microcontroller that controls and determines the pattern and session of the display. The implementation of a moving message displays a text containing 22 characters (i.e. WELCOME TO DEPT. OF PHYSICS). The electronic message display helps to pass information, educates, enlightens, facilitates commercial activities through advertisement and marketing of goods and services, description of places, etc The ease in which it displays information makes it a veritable, suitable, and an excellent tool for passing information fast and pleasurable to the public. Furthermore, it enhances the response to information in an attractive way and manner in which it displays messages. The microcontroller used in this work is the PIC16F84A. It belongs to a class of 8-bit microcontrollers of RISC (reduced instructions set) architecture. Its output controls the switching of the relays through a transistor switching stage that switches its socket. The LED matrix (array) is arranged in parallel and soldered to a Vero board with the microcontroller and other electronic components. Such as resistor, capacitors, transistor, relays, LEDs, diodes, transformer.

## 1.0 Introduction

The role of information dissemination in the society cannot be overemphasized, because it enlightens, educates, entertains, and facilitates commercial activities through advertisement and marketing of goods and services. Its uses in surveillance and monitoring (security, traffic control etc) and description of places remain indispensable. Various means of information dissemination include broadcasting (radio and television), the internet, newspaper, bill boards, sign posts and neon displays etc. The choice of any information display garget depends on factors such as the target audience.

However, regardless of the medium employed, the most important thing is the fulfillment of the purpose for which the information is disseminated. The information should be relevant to the intended audience with clarity, objectiveness and a good degree of comprehension. In addition, response to information is further enhanced when the medium that is employed is attractive and highly inviting, as in the case of a moving message display.

A moving message display could also be referred to as an electronic notice board. It is an electronic gadget that is capable of displaying lengths of alphanumeric characters, symbols and any form of representation on a matrix (array) of light emitting diodes (LEDs).It scrolls and moves the character across the screen making the emerging information readable and appealing to the audience. It also has an unparalleled advantage in drawing people's attention by causing them to reflect many times on the scrolling lights automatically displaying either messages of advertisement, place description or greetings at any time of the day. It can be used for both indoor and outdoor purposes. In fact it is the most alluring, unique captivating and attractive means of information dissemination.

The moving message display can be realized in different ways with operational possibilities achievable varying accordingly with the design. This design features a USB keyboard user interface with which the user can input or edit data to be displayed.

The research relating to the design of LED displays and microcontroller-based circuits was investigated throughout this paper. This knowledge provides better methods for approaching the task while helping to gain an understanding of microcontroller systems that will be useful for future works. The design for a sign message display that can be mounted outdoor the department, is then implemented in this work in the initial prototype, to verify its physical viability and to investigate any possible improvements. From which a viable product can be developed in the future. Hence, this paper aims to design and construct a microcontroller based moving message display with the following objectives/scope:

- Design a microcontroller-based circuits and size battery that are sufficient to power the LED moving message displays panel.
- Study the specifications of all the display system components (data sheet).
- Design the needed software that can provide the pre-programmed messages, and that of the animations for the display panel.
- Construct a prototype of the display system and studies its performance.

## 2.0 Materials and methods

### 2.1 Model with the incorporation of a microprocessor

Microprocessor was bused in the design of digital billboards, with some other support components, the LED was arranged in matrix form and an EPROM then programmed to hold the instruction and data, for the processor to manipulate. The model is shown in Figure 2.1. The various constituents that make it functional include:

- The microprocessor which is the central processing unit
- The memory which invariably comprises both read or write and read only devices (RAM and ROM respectively).
- Interface devices to facilitate input and output (I/O) for peripheral devices like display unit which in this case is LED matrix.

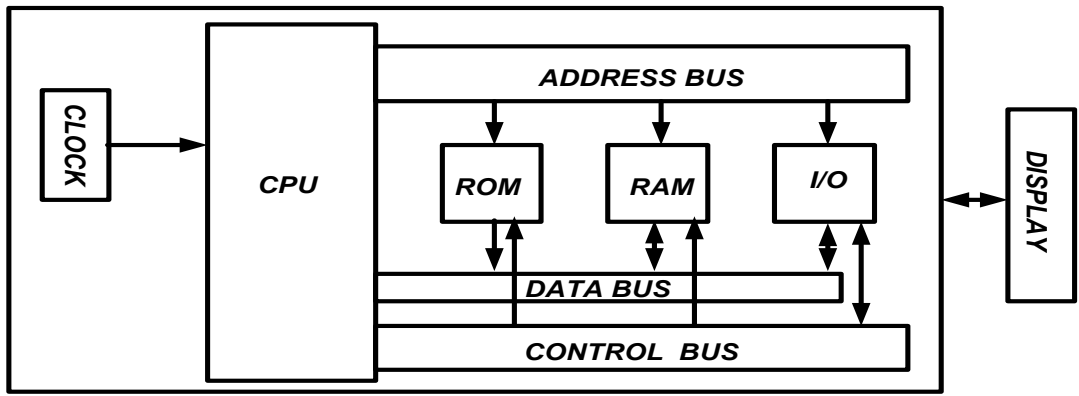

**Figure 2.1**: Block diagram of Moving Message Display Board designed with the incorporation of processor.

### 2.2 Model that incorporates both processor and keyboard

With the incorporation of microprocessor and other peripheral devices like keyboard in the design of digital billboard, the task of reprogramming for different messages is eliminated. This is as a result of the possibility of entering at will, the various characters constituting any message desired for display through the use of a keyboard rather than having to go through the more rigorous process of erasing the contents of the EPROM for re-programming before new messages could be displayed.

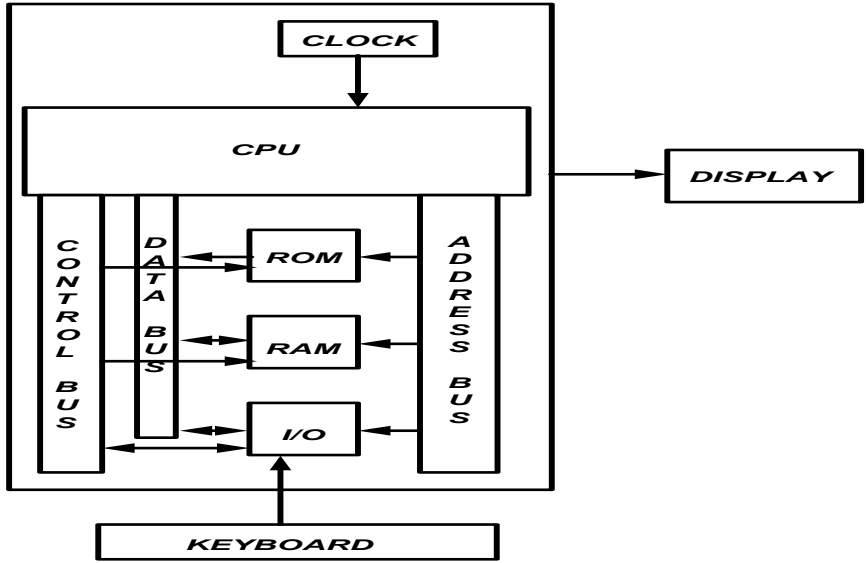

**Figure 2.2:** Block diagram of Moving Message Display Board designed with the incorporation of processor and a keyboard.

## 2.3 Model which incorporates a microcontroller and keyboard

The design of a microcontroller-based moving message display with keyboard is similar to based-microprocessor. The major differences is mainly in the reduction in hardware complexity which the use of a micro controller offers due to its non-requirement of many external support chips such as timers, memory, input/output devices etc. As well as the flexibility offered with the included keyboard as an input device and other possibilities achievable in the microprocessor-based display are not left out in the model.

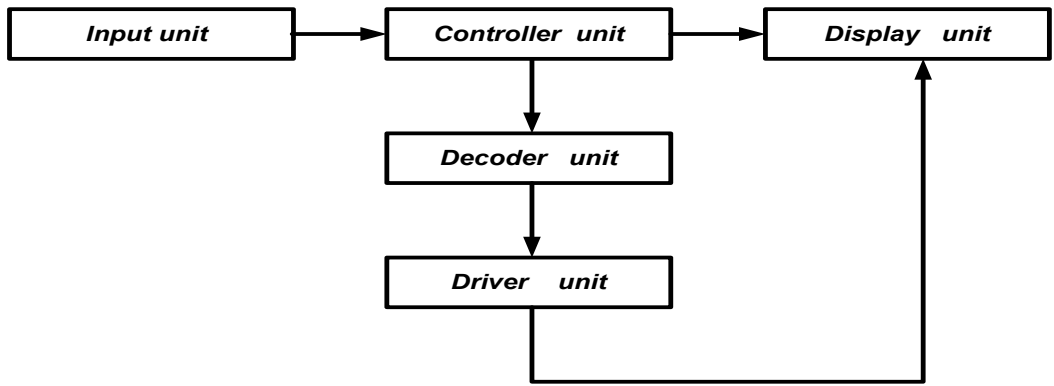

**Figure 2.3:** Model which incorporates a microcontroller and keyboard

## 2.4 Description of basic components constituting the system model

i. **IC voltage regulator**

The LM78XX series of three terminal regulators several fixed output voltages which make them useful in a wide range of applications. One of these is local on card regulation, eliminating the distribution problems associated with single point regulation. The voltages available allow these regulators to be used in logic systems, instrumentation, HiFi, and other solid state electronic equipment. Although designed primarily as fixed voltage regulators, these devices can be used with external components to obtain adjustable voltages and currents.

The LM78XX series is available in an aluminum package which will allow over 1.0A load current if adequate heat sinking is provided. The current limiting is included as well to limit the peak output current to a safe value. Safe area protection for the output transistor is also provided to limit internal power dissipation.

7805 IC regulator is used in this project to maintain a stable +5 Volt required to drive the microcontroller. Figure 2.6 shows the external look of 7805 IC Regulator.

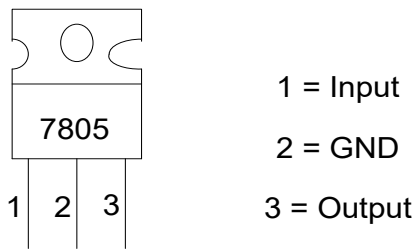

1 = Input
2 = GND
3 = Output

**Figure 2.4**: The external look of 7805 IC Regulator.

ii. **Microcontroller**

A microcontroller is a computer control system on a single chip. It has many electronic circuits which can decode written instructions and convert them into electrical signals. The microcontroller also goes further, stepping through the instructions and executing them one by one.

**Constituents of a Microcontroller**

A microcontroller includes EPROM program memory, user RAM for storing data, timer circuits, an instruction set, special function registers, power-on reset, interrupts, low power consumption and a security bit for software protection.

**The Basic Microcontroller System**

The basic microcontroller system is explained in the block diagram below

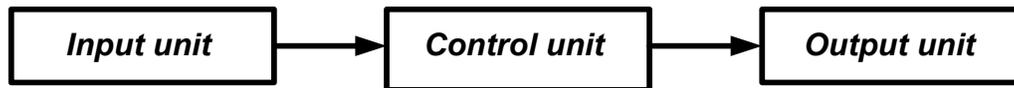

**Figure 2.5**: the block diagram for the basic microcontroller system

**iii, The Input Unit:** The input component consists of digital devices such as switches, push buttons, pressure mats, float switches, keypads, and radio receivers' e.t.c. It also consists of analogue sensors such as light dependent resistors, thermistors, gas sensors, pressure sensors, e.t.c.

iv. **The Control Unit:** This is of course the microcontroller itself. The microcontroller monitors the inputs and as a result of the program written into it, turns output ON and OFF. The microcontroller stores the program in its memory and executes the instructions under the control of the clock circuit.

v. **The Output Unit:** This consist of output devices such as light emitting diodes, buzzers, motors, alphanumeric displays, radio transmitters, 7-segment displays, e.t.c. Microcontroller is similar to microprocessor but with addition of I/O ports, memory, counter, a clock and interrupt circuitry. It is the additional circuitry that makes the microcontroller such a unique device. The microcontroller is designed primarily to operate on data that is fetched through serial or parallel input ports. The data is operated on, under the control of software stored in ROM and external device controlled though signal, fed via the output port. The incoming data may be from receiving and operating on the data can be established using the interrupt control circuitry.

Microcontroller is designed to operate with the minimum of external circuitry to perform control-oriented task using a control programme ROM. The instruction set for the microcontroller is simpler than that of the microprocessor. Most of its instructions will move code and data from internal memory to Arithmetic and Logic Unit (ALU). Also, the use of many inputs/output pins allow data to be moved between file microcontroller and external devices as single bits. The operation on single bit such as logical operation, flag settings/clearing, is unique to the microcontroller. Microcontrollers are available in various sizes. Such as 8-bit and 16-bit, the 16-bit is popular for higher performance specifications.

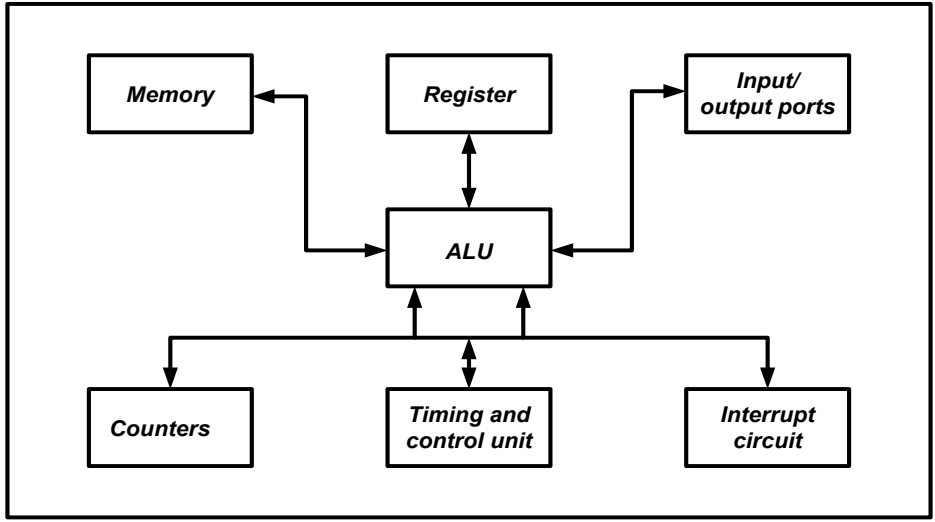

**Figure 2.6**: the basic arrangement of the internal structure of the microcontroller.

The ALU is the Arithmetic and Logical Unit which performs arithmetic manipulations such as binary additions, subtractions and possibly multiplications and divisions. Also logic functions such as AND, OR, NOT and Ex-OR can be implemented. The ALU consists of gates, which are organized to receive binary inputs and provides binary output according to the instruction codes. The register group contains the data that the processor needs while performing the task of executing a program. The registers include the Program Counter (PC), the Accumulator and Stack Pointer.

**The Input/output Unit**

This system requires an interface with "outside world". The input/output interface allows the connection of input data through a keyboard and sensors which can transpose information such as movement of words or pictures into electrical signal. For output data these could be a monitor for displaying instructions, data and output that can feed external devices such as relays, LED (Light Emitting Diode). Without means of input and output, a microcontroller system will be of little use because there would be no means of communication with each other.

The amount of I/O provided in any particular microcontroller system is of course largely determined by the range of applications envisaged. In this design for example we expected to accept input from the keyboard during the data programming and display output during running programming respectively. Other components such as resistors, capacitors etc. are used

**The power supply unit**

This is the unit from which the maximum of 5 V DC voltage ever needed in the circuit is being supplied. The unit contains the following components

- ❖ A transformer: This step down the a.c supply voltage to suit the requirement of the solid-state electronic devices. It also provides isolation from the supply line -
- ❖ A full wave bridge rectifier: This performs the transformation of the a.c. voltage into a pulsating d.c. voltage in a process known as rectification
- ❖ Capacitors: These filter out (remove) fluctuations or pulsations (called ripples) present in the rectifier output voltage.

A fixed 1C voltage regulator (78LM05) is mounted on the internal circuit board to provide a stable + 5V d.c. output from the unregulated 12V d.c. fed into its input from the a.c/d.c. adaptor.

**Calculations of value(s)**

Using a 240V transformer on a 50Hz supply and transformer secondary *r.m.s* voltage output is 12V.

*Peak voltage*, $V_p = V_{rms} \times \sqrt{2}$

$V_p = 12\sqrt{2} = 16.97V$

*Supply frequency*, $f = \dfrac{1}{period(T)} = 50Hz$

*Period*, $T = \dfrac{1}{f} = \dfrac{1}{50} = 0.02s = 20ms$

The total voltage drop, $V_d$, for the two diodes involved in the rectification process in either of positive or negative cycles,

$V_d = 2V_{BE}$ [$V_{BE} = 0.7V$ for a silicon diode]

$V_d = 2 \times 0.7V = 1.4V$

Actual peak voltage value, $V_{LM} = (V_m - 2V_{BE})V$

$V_{LM} = (16.97 - 1.4)V$

$V_{LM} = 15.57V$

Change in peak voltage value over the discharge period, $\delta V = V_{LM} - V_{dc}$

$V_{dc} = 10V$

The filter capacitor should not discharge down to 6V in accordance with the input voltage specification of the voltage regulator.

$\delta V = (15.57 - 10.0) = 5.57V$

Change in time over the discharge period, $\delta t = 10ms$

Total current consumption for this design is not expected to exceed 600mA

Hence the value of the filter capacitor is obtained thus:

$$C = \frac{600mA \times 10ms}{5.57V} = 1077.20\mu F$$

To provide a safety margin, the capacitor value chosen is twice the calculated value which implies a value 2154.4µF.

The nearest available capacitor value of 2,200µF is used as the filter capacitor.

**Table 1.1: PIC 16F628A Specification**

| Device | Program Memory | EEPROM data memory (Bytes) | RAM | I/O | 10-bit AD channels | |
|---|---|---|---|---|---|---|
| 16F628 | 8K | 156 | 368 | 13 | 33 | 8 |

**Microcontroller Specification**

Various factors are considered in the choice of microcontroller to use for a particular purpose.
In this paper, of this kind where the number of character that can be displayed is largely dependent on the amount of memory available, a microcontroller with a large memory sufficient input/output ports and analogue/digital channels such as the PIC 16F628A is used.

**PIC16F628A**

The microcontroller unit used in this paper is the PIC16F84A and it belongs to a class of 8-bit microcontrollers of RISC (reduced instruction set) architecture. It is an 18 pin dual in-line package chip. The PIC is a tiny but complete computer. It has a CPU (central processing unit), program memory (PROM), working memory (RAM), and two input-ports. The CPU is the "brain" of the computer. It reads and executes instructions from the program memory. As it does so, it can store and retrieve data in working memory (RAM). CPUs make a distinction between "registers" located within the CPU and "RAM" outside it; the PIC does not, and its general-purpose working RAM is also known as registers." On the 'F84, there are 68 bytes of general-purpose RAM, located at addresses C to hex 4F.

Besides the general-purpose memory, there is a special "working register" or register" where the CPU holds the data it is working on. There are also several special function registers each of which controls the operation of the PIC in some way. The program memory of the F84A consists of flash EPROM; which can be recorded and erased electrically.Itretains its contents when powered off. Program memory (FLASH) is used for storing a written program. Since memory made in FLASH technology can be programmed and cleared more than once, it makes this microcontroller suitable for device development. EEPROM - data memory that needs to be saved when there is no supply. It is usually used for storing important and must not be lost if power supply suddenly stops.   Many other PICs require ultraviolet light for erasure and are not erasable if you buy the cheaper version without the quartz window. The chip, however, is always erasable and reprogrammable.

There are two input- output ports, port A and port B, and each pin of each port can be set individually as an input or an output. The bits of each port are numbered, starting from 0. In output mode, bit 4 of port A has an open collector (or rather open drain); the rest of the outputs are regular CMOS. The CPU treats each port as one 8-bit byte of data even though only five bits of port A are actually brought out as pins of the IC. PIC inputs are CMOS-compatible; PIC outputs can drive TTL or CMOS logic chips: Each output pin can source or sink $20mA$ as long as only one pin is doing so at a time

**Pin Description**

IC16F628 has a total of 18 pins. It is most frequently found in a DIP18 type of case but can also be found in SMD case which is smaller from a DIP. DIP is an abbreviation for Dual in Package. SMD is an abbreviation for Surface Mount Devices suggesting that holes for pins to go through when mounting, are not necessary in soldering this type of a component.

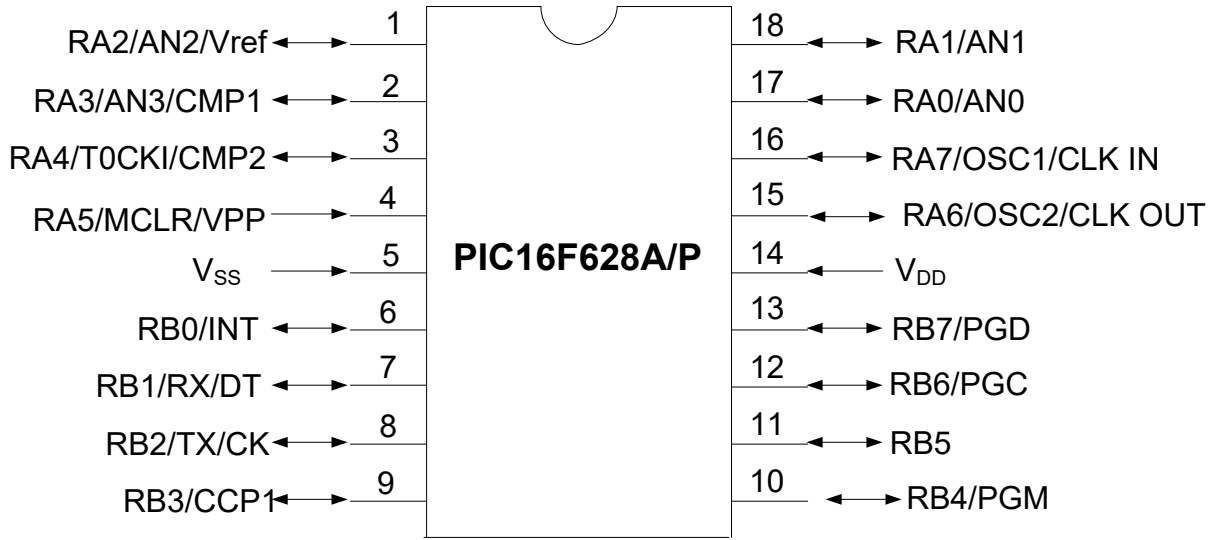

**Figure. 2.9**: Top View of The PIC16F84A Showing the Pin out Orientation of the IC.

Pins on P1C16F628A microcontroller have the following meaning.

Pin no. 1-*RA2* Second pin on port A. Has no additional function.

Pin no.2-*RA3* Third pin on port A. Has no additional function.

Pin no.3-*RA4* Fourth pin on port A. TOCK1 which functions as a timer is also found on this pin.

Pin no.4- MCLR Reset input and $V_{pp}$ programming voltage of a microcontroller.

Pin no.5-$V_{ss}$ Ground of power supply.

Pin no. 6- RB0 Zero pin on port B. Interrupt input is an additional function.

Pin no. **7-** RBI First pin on port B. No additional function.

Pin no.8- RB2 Second pin on port B. No additional function.

Pin no. 9- RB3 Third pin on port B. No additional function.

Pin no. 10- RB4 Fourth pin on port B. No additional function.

Pin no. 11- RB5 Fifth pin on port B. No additional function.

Pin no. 12- RB6 Sixth pin on port B. 'Clock' line in program mode.

Pin no. 13- RBI Seventh pin on port B. 'Data' line in program mode.

Pin no. 14-$V_{dd}$ Positive power supply pole.

Pin no. 15- OSC2 Pin assigned for connecting with an oscillator.

Pin no. l6- OSC1 Pin assigned for connecting with an oscillator.

Pin no. 17- RA2 Second pin on port A. No additional function.

Pin no.18- RA1 First pin on port A. No additional function.

**Power and Clock Requirements**

The PIC16F628A requires a 5-Volt supply, (4.0 to 6.0 volts) will be okay for IC, so it can be run from three 1.5-Volt cells. The PIC consumes only 1$m$A-even less, at low clock speeds - but the power supply must also provide the current flowing through LEDs or other high-current devices that the PIC may be driving.

Like any CPU, the PIC needs a clock - an oscillator to control the speed of the CPU and step it through its operations. The maximum clock frequency of the PIC16F84A is as 4MHz. There is no lower limit. Low clock frequencies save power and reduce the amount of counting the PIC has to do when timing a slow operation. At 30 kHz, a PIC can run on $0.1mA$.

The clock signal can be fed in from an external source, or the PIC's on-board oscillator with either a crystal or a resistor and capacitor can be used. Crystals are preferred for high accuracy e.g. 3.58-MHz crystals. The resistor-capacitor oscillator is cheaper but not used in time dependent operations.

**RAM –Data Memory**

This is used by a program during its execution. In RAM, data is stored all inter-results or temporary data during run-time. PORTA and PORTB are physical connections between the microcontroller and the outside world. Port A has five, and port B has eight pins. FREE-RUN TIMER is an 8-bit register inside a microcontroller that works independently of the program. On every fourth clock of the oscillator it increases its value until it reaches the maximum and then it starts counting over again from zero. As we know the exact timing between each two increments of the timer contents, timer can be used for measuring time which is very useful with some devices.

**Central Processing Unit**

This has a role of connecting element between other blocks in the microcontroller. It coordinates the work of other blocks and executes the user program. In system programmability of this chip (along with using only two pins in data transfer) makes possible the flexibility of a product, after assembling and testing have been completed. This capability can be used to create assembly-line production, to store calibration data available only after final testing, or it can be used to improve programs on finished products.

**Counter/Timer**

There is a internal 8 bit counter/timer that sets a flag when it rolls over from 255 to zero. This can be used as a counter or timer. As a timer, it is connected to the internal clock and increments at the clock frequency divided by four. A flag can be polled to tell when time is up. The timer can also be set up to generate an interrupt when this happens. It wouldn't take long to count all the way up at $1\ MHz$ so a programmable 'prescaler' can be used. The prescaler can be set to give output pulses at ratios of 1:2, 1:4, 1:8 etc. up to 1:256, extending the timeout up to the tens of millisecnds range for a $4\ MHz$ clock.

**Transistor switching stage**

The output of the microcontroller controls the switching of the relays via the transistor switching stages, which switches the socket. The power of the load to be activated depends on the relay contact rating; hence the choice of relay is critical.

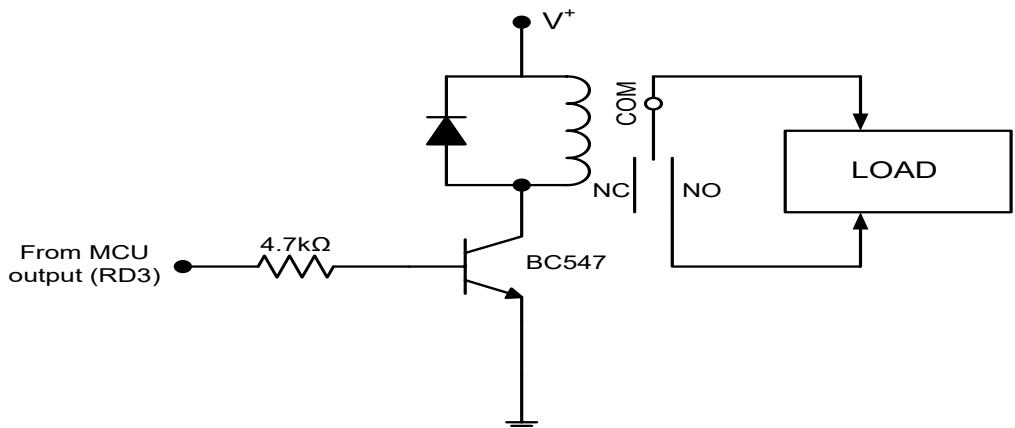

**Figure** 2.0 : Switching transistor stage.

**The driver unit**

This unit is made up of NPN transistors which drive the particular LED(s) to be turned on through the supply of an active low at the cathode of the LED concerned. These are the base resistors for each of the NPN transistors driving the LED matrix. The 1k value used is justified thus.

Total current consumption per LEDs arrangement is given by current consumption per LED x total number of LEDs.

The display unit is designed to consist of five display modules with each letter
consisting of about 20 LEDs matrix, i.e. 444 LEDS altogether.

    The anodes of all LEDS in the same row for each board are connected together

while the cathodes of all LEDS in the same column are likewise connected together. The lines connecting the anodes on each of the rows constitute the date lines while the lines connecting the cathodes of the LEDS on each of the columns are the address lines.

The 150ohm resistors are serving as current-limiting devices to prevent the connected LEDs from drawing excessive current and getting destroyed. The 150 value used is justified thus;

Since the maximum current to power an LED is 30mA and the needed voltage being between 1.5 to 3.3V, then using safe values of 20mA and 3V,

**Display Scanning**

A commonly used method to control the illuminated displays is to turn the rows or columns of the display on and off in quick succession. Multiplexing, as it is termed, reduces the amount of input and output (I/O) lines required to control the individual elements of a large display. In multiplexing a common set of control lines is used to join each display to the control system. As a consequence the amount of conductors and ports required for controlling is significantly reduced when compared with connecting each display individually to the system.The principle of scanning is employed in the turning ON of the LEDs for character display by scanning of each character. To put ON a particular LED in the matrix, its row is brought HIGH and the column LOW. This is achieved with the aid of drivers which drive the columns of the particular LED to be turned ON through the supply of an active LOW at the LED cathode as an inverted form of the active HIGH logic output of the decoder.

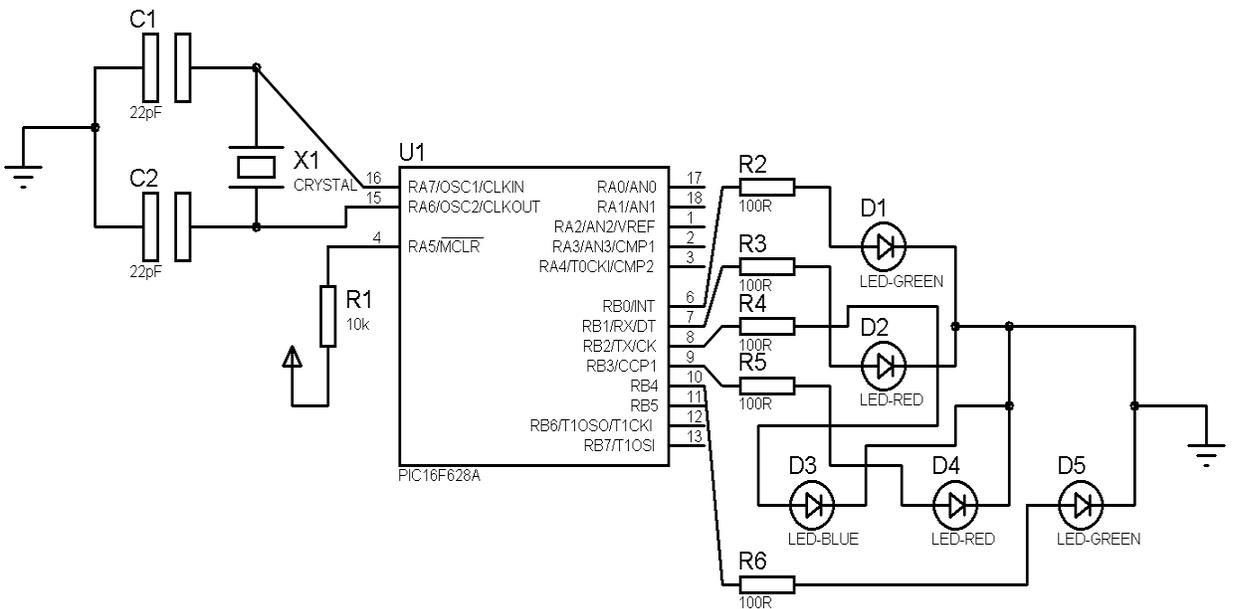

**Figure** : Circuit diagram of the moving message display board microcontroller

 SOURCE CODE OF THE PROGRAM FOR THE DISPLAY

/*Running LED message display using PIC16F628A

@4MHz */

//LED configuration

sbit WELCOME at RB0_bit;

sbit TO at RB1_bit;

sbit DEPT at RB2_bit;

sbit OF at RB3_bit;

sbit PHYSICS at RB4_bit;

void main() {

TRISA=0;

TRISB=0;

PORTA=0;

PORTB=0;

    do{

```
PHYSICS=0;
WELCOME=1;
delay_ms(500);
WELCOME=0;
TO=1;
delay_ms(500);
TO=0;
DEPT=1;
delay_ms(500);
DEPT=0;
OF=1;
delay_ms(500);
OF=0;
PHYSICS=1;
delay_ms(500);
PHYSICS=0;
WELCOME=1;
delay_ms(500);
WELCOME=0;
TO=1;
delay_ms(500);
TO=0;
DEPT=1;
delay_ms(500);
DEPT=0;
OF=1;
delay_ms(500);
OF=0;
PHYSICS=1;
delay_ms(500);
PHYSICS=0;
```

```
WELCOME=1;
delay_ms(500);
WELCOME=0;
TO=1;
delay_ms(500);
TO=0;
DEPT=1;
delay_ms(500);
DEPT=0;
OF=1;
delay_ms(500);
OF=0;
PHYSICS=1;
delay_ms(500);
PHYSICS=0;
WELCOME=1;
delay_ms(500);
WELCOME=0;
TO=1;
delay_ms(500);
TO=0;
DEPT=1;
delay_ms(500);
DEPT=0;
OF=1;
delay_ms(500);
OF=0;
PHYSICS=1;
delay_ms(500);
PHYSICS=0;
delay_ms(1500);
```

```
WELCOME=TO=DEPT=OF=PHYSICS=1;
delay_ms(800);
WELCOME=TO=DEPT=OF=PHYSICS=0;
delay_ms(800);
WELCOME=TO=DEPT=OF=PHYSICS=1;
delay_ms(800);
WELCOME=TO=DEPT=OF=PHYSICS=0;
delay_ms(800);
WELCOME=TO=DEPT=OF=PHYSICS=1;
delay_ms(800);
WELCOME=TO=DEPT=OF=PHYSICS=0;
delay_ms(800);
WELCOME=TO=DEPT=OF=PHYSICS=1;
delay_ms(800);
WELCOME=TO=DEPT=OF=PHYSICS=0;
delay_ms(1500);
WELCOME=1;
delay_ms(400);
WELCOME=0;
delay_ms(400);
WELCOME=1;
delay_ms(400);
WELCOME=0;
delay_ms(400);
WELCOME=1;
delay_ms(400);
WELCOME=0;
delay_ms(400);
WELCOME=1;
delay_ms(400);
WELCOME=0;
```

```
delay_ms(400);
PHYSICS=0;
TO=1;
delay_ms(500);
TO=0;
DEPT=1;
delay_ms(500);
DEPT=0;
OF=1;
delay_ms(500);
OF=0;
PHYSICS=1;
delay_ms(500);
PHYSICS=0;
WELCOME=1;
delay_ms(400);
WELCOME=0;
delay_ms(400);
WELCOME=1;
delay_ms(400);
WELCOME=0;
delay_ms(400);
WELCOME=1;
delay_ms(400);
WELCOME=0;
delay_ms(400);
WELCOME=1;
delay_ms(400);
WELCOME=0;
delay_ms(400);
PHYSICS=0;
```

```
TO=1;
delay_ms(500);
TO=0;
DEPT=1;
delay_ms(500);
DEPT=0;
OF=1;
delay_ms(500);
OF=0;
PHYSICS=1;
delay_ms(500);
PHYSICS=0;
WELCOME=1;
delay_ms(400);
WELCOME=0;
delay_ms(400);
WELCOME=1;
delay_ms(400);
WELCOME=0;
delay_ms(400);
WELCOME=1;
delay_ms(400);
WELCOME=0;
delay_ms(400);
WELCOME=1;
delay_ms(400);
WELCOME=0;
delay_ms(400);
PHYSICS=0;
TO=1;
delay_ms(500);
```

```
TO=0;
DEPT=1;
delay_ms(500);
DEPT=0;
OF=1;
delay_ms(500);
OF=0;
PHYSICS=1;
delay_ms(500);
PHYSICS=0;
delay_ms(1500);
WELCOME=TO=DEPT=OF=PHYSICS=1;
delay_ms(5000);
WELCOME=TO=DEPT=OF=PHYSICS=0;
delay_ms(1500);
WELCOME=TO=1;
delay_ms(500);
WELCOME=TO=0;
delay_ms(500);
WELCOME=TO=1;
delay_ms(500);
WELCOME=TO=0;
delay_ms(500);
WELCOME=TO=1;
delay_ms(500);
WELCOME=TO=0;
delay_ms(500);
DEPT=OF=PHYSICS=1;
delay_ms(8000);
DEPT=OF=PHYSICS=0;
WELCOME=TO=1;
```

```
delay_ms(400);
WELCOME=TO=0;
delay_ms(400);
PHYSICS=1;
delay_ms(400);
PHYSICS=0;
delay_ms(400);
DEPT=1;
delay_ms(400);
DEPT=0;
delay_ms(400);
WELCOME=TO=1;
delay_ms(400);
WELCOME=TO=0;
delay_ms(400);
PHYSICS=1;
delay_ms(400);
PHYSICS=0;
delay_ms(400);
DEPT=1;
delay_ms(400);
DEPT=0;
delay_ms(400);
WELCOME=TO=1;
delay_ms(400);
WELCOME=TO=0;
delay_ms(400);
PHYSICS=1;
delay_ms(400);
PHYSICS=0;
delay_ms(400);
```

```
    DEPT=1;
    delay_ms(400);
    DEPT=0;
    delay_ms(6000);
    }while(1);}
```

## 2.6 Construction

The physical realization of the work is very vital. Here the paper work is transformed into a finished hardware.

**Implementation**

The implementation of this project was done on the breadboard. The power supply was first derived from a bench power supply in the school electronics lab. To confirm the workability of the circuits before the power supply stage was soldered. The implementation of the project on bread board was successful and it met the desired design aims with each stage performing as designed.

**Soldering**

The various circuits and stages of this work were soldered in tandem to meet desired workability of the project. The LED matrix array stage was first soldered before the microcontroller, shift register and switching stages were done. The soldering of the project was done on Vero- board, and was soldered on a Vero board. The Vero board contains the power supply stage, the microcontroller stage with the shift register stage.

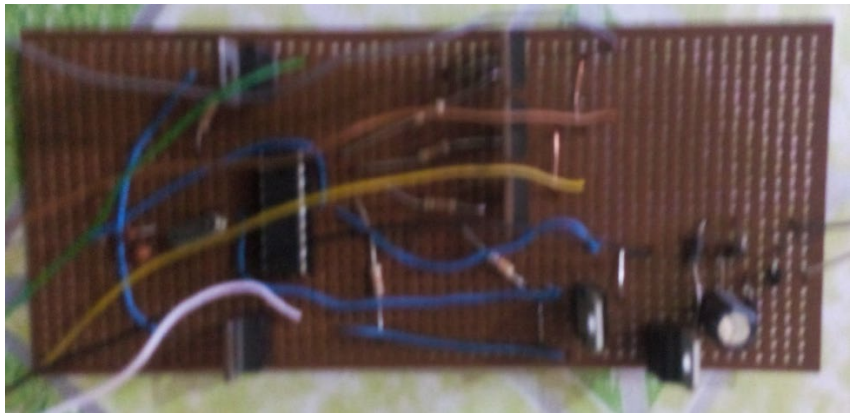

Figure 4.1a components layout on Vero-board 1.

**Casing and Boxing.**

The third phase of the work is the construction of casing /boxing. The casing material being plastic designed with special perforation and vents were well labeled to give ecstatic value.

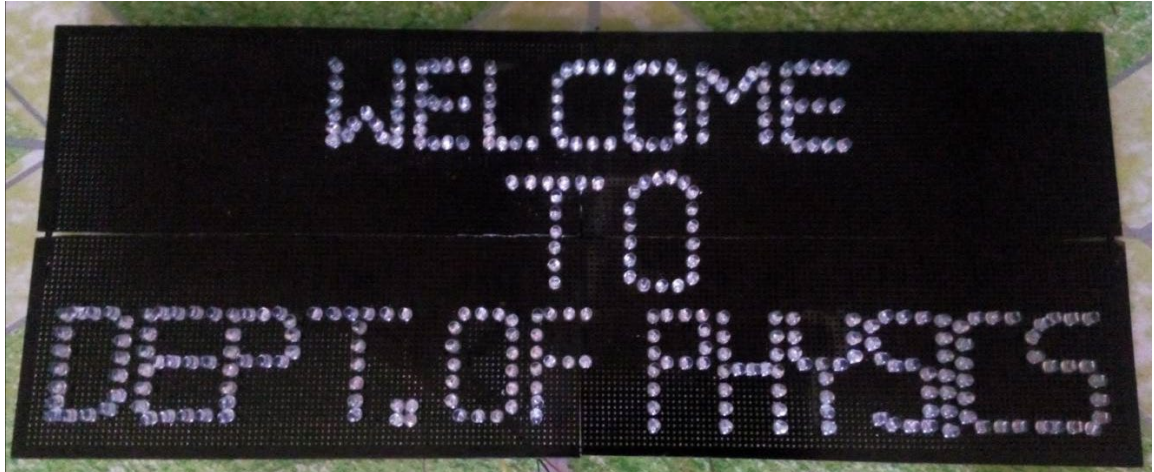
**Figure .2.2**: Prototype before being cased.

**Testing**

Stage by stage testing was done according to the block representation on the breadboard, before soldering of circuit commenced on Vero board.

The process of testing and implementation involved the use of some test and measuring equipments stated below.

1. **Bench Power Supply**: This was used to supply voltage to the various stages of the circuit during the breadboard test before the power supply in the project was soldered. Also during the soldering of the project the power supply was still used to test various stages before they were finally soldered.
2. **Oscilloscope**: The oscilloscope was used to observe the ripples in the power supply waveform and to ensure that all waveforms were correct and their frequencies accurate. The waveform of the oscillation of the crystal oscillator used was monitor to ensure proper oscillation at 4MHz.
3. **Digital Multi-meter**: The digital multi-meter basically measures voltage, resistance, continuity, current, frequency, temperature and transistor $h_{fe}$. The process of implementation of the design on the board required the measurement of parameters like, voltage, continuity, current and resistance values of the components and in some cases frequency measurement. The digital multimeter was used to check the output of the voltage regulators used in this project.

**Conclusion**

The set goals within this paper have been achieved, strictly carried out the objectives, proving the feasibility of the display. The prototype was extremely developed economically using several boards on which the components were mounted. Through this type of construction, modifications could be easily made before entering into the development of larger scales, and more costly prototypes. Other factors such as availability of components efficiency, compatibility, portability and durability were considered. The performance of the device after test met design specifications. The work was quite challenging and tedious but eventually was a success.